\documentclass[12pt]{article}
\usepackage{cite}
\usepackage{graphicx}
\usepackage{amsmath}
\usepackage[byname]{smartref}
\usepackage{makeidx}
\usepackage{imakeidx}
\makeindex[columns=1]
\usepackage[bookmarks,pagebackref, pdfpagelabels=true, colorlinks=true, urlcolor=blue, citecolor = blue, anchorcolor = blue, linkcolor=blue, pdfborder={0 0 0}]{hyperref}
\usepackage{txfonts}
\usepackage{tocloft}
\usepackage{calrsfs}

\usepackage{algorithm}
\usepackage{algpseudocode}
\usepackage{adjustbox}
\usepackage{xspace}
\usepackage{mathrsfs}
\usepackage{breqn}
\usepackage{amsfonts, amssymb}
\usepackage{tikz}
\usepackage{tikzscale}
\usepackage{textcomp}
\usepackage{xcolor}
\usepackage{color}
\usepackage{sidecap}
\usepackage{graphicx}
\usepackage{pdfpages}
\usepackage{lipsum}
\usepackage{subcaption}

{\clearemptydoublepage 
 \begin{center}
  \section*{Dedication}
 \end{center}
 \begingroup
}{\newpage\endgroup}

{\pagestyle{plain}\pagenumbering{roman}}%
{\pagenumbering{arabic}}

\newcommand{\qed}{\nobreak \ifvmode \relax \else
      \ifdim\lastskip<1.5em \hskip-\lastskip
      \hskip1.5em plus0em minus0.5em \fi \nobreak
      \vrule height0.75em width0.5em depth0.25em\fi}

\usepackage{color}
\newboolean{showComments}
\newcommand{\todocmd}[1]{\small{\textcolor{red}{#1}}}

\newcommand{\todo}[1]{\ifthenelse {\boolean{showComments}} {\todocmd{#1}} {}}

\newcommand{\Rdef}{\mathcal{R}}
\newcommand{\tol}{\mathbb{T}}
\numberwithin{equation}{section}

\newcommand{\maple}[1]{{\tt #1}\xspace}

\makeatother

\begin{document}


\title{Contributions to the Algorithmic Foundations of Approximate Lie Symmetry Algebras of Differential Equations}


\author{Siyuan Deng\\
\textit{Department of Mathematics} \\
    \textit{University of Western Ontario}\\
    London, Canada \\
    sdeng53@uwo.ca \\
    \and
Gregory Reid\\
\textit{Department of Mathematics} \\
    \textit{University of Western Ontario}\\
    London, Canada \\
    reid@uwo.ca}

\date{}

\maketitle

\begin{abstract}

Symmetry transformations that leave an object invariant, play a fundamental role in science and mathematics.  In this article we consider Lie symmetry transformations, that depend analytically on their parameters, and leave the solution set of differential equations invariant.  Such Lie symmetry groups uniquely determine their Lie symmetry algebras, which are the linearized form of the transformations near the identity transformation.  Given a differential equation model, which is polynomially nonlinear in its dependent variable and their derivatives, and a rational function of its independent variable, exact differential elimination algorithms have been developed that determine the dimension and structure constants of the Lie symmetry algebra of the differential equation.

However, models arising in applications often have approximate parameters, and direct application of these symbolic algorithms above is prone to instability since these algorithms strongly depend on the orderings of the variables involved, much like the instability encountered by naive Gaussian elimination.  This motivates the need to address questions at the algorithmic foundation of approximate Lie symmetry algebras of differential equations.
How do we define approximate Lie symmetry? How do we compute and apply approximate Lie symmetry algebras of differential equations?  How reliable are the results?

To address such questions, we define approximate Lie symmetry algebras in terms of exact Lie symmetry algebras of a nearby differential equation. Our algorithm for identifying these hidden approximate Lie symmetry algebras uses the SVD to find nearby rank-deficient systems combined with approximate geometric involutive forms of the approximate symmetry-defining systems. Our approach is local and depends on an input base point in the space of independent and dependent variables. 

For those exact rational differential equations where the exact differential equations provide a unique result, they appear in our experiments as very close by approximate Lie algebras.  In general, our local approach can yield many different nearby Lie symmetry algebras.  We also outline a numerical approach to determining the approximate isomorphic of such Lie symmetry algebras as the local base point varies across a grid in the base space. We also give a method for evaluating the reliablity of our results. This enables the base space to be partitioned into regions where different local Lie symmetry algebras are admitted, separated by unstable transition regions.

\end{abstract}

\section{Introduction}
\label{sec:chap4: intro}

The foundation is laid for developing approximate Lie symmetry algebras for differential equations (DEs), which extend classical exact Lie symmetry theory to accommodate systems with approximate or imprecise coefficients. The need for approximate symmetries arises in practical applications where small errors or perturbations in the equations prevent the construction of exact symmetries. By employing numerical methods such as singular value decomposition (SVD), this approach allows for the analysis of systems that exhibit near-symmetries, facilitating the study of both exact and approximate components within the same framework.

Building on this foundation, we introduce a novel method for estimating the quality of the derived approximate Lie symmetry algebras. This method is based on finding the structure constants of the Lie algebra and evaluates the accuracy of the symmetries by estimating the relative deviation from the commutators lying in the span of the Lie symmetry operators.



Given a differential equation or system of differential equations $R$, an one ($\epsilon$) parameter Lie point symmetry is an analytic transformation of that preserves the form of $R$.  The corresponding Lie algebra, is the linearization at the identity of the Lie group.  Multi-parameter Lie groups can be obtained via the one parameter case.  In the exact case, where $R$ is polynomially nonlinear with rational coefficients, symbolic algorithms have been developed that produce and reduce their Lie algebra defining systems to standard forms (or differential bases), from which the structure of the Lie algebra can be determined.  In this case, one may in theory analyze entire classes of differential equations (e.g. second order ODE) for their Lie symmetry algebras.

But in practical applications, one may have a differential equation which is not a differential polynomial with rational coefficients.  For example, $R$ may not be a differential polynomial or $R$ may involve approximate parameters.  Na\"ive application of the exact differential-elimination algorithms to the Lie algebra defining systems, is prone to instability (in the same way that na\"ive Gaussian elimination is unstable).

Recent progress has been made in the development of approximate differential-elimination algorithms for such approximate systems. In this article, we use an algorithm based on the SVD of the coefficient matrices appearing in the computations.
That algorithm has been used previously with encouraging preliminary results but without a clear definition of approximate symmetries. 

We now describe several approaches to determining approximate symmetries, including one that uses exact symmetries followed by approximate fitting analysis.  A well-known example is the Harmonic Oscillator Schr\"{o}dinger equation, which admits a rich 
symmetry algebra \cite{dekker1981classical}.  
Then solutions of the Schr\"{o}dinger equations with potentials having a local minimum are approximated using symmetry invariant solutions of 
a Harmonic Oscillator Schr\"{o}dinger equation in a neighborhood of the minimum.

From the algorithmic point of view this approach can be expanded to the case algorithm classification of exact symmetry algebras of classes of differential equations followed by approximate fitting analysis (algorithmic exact symmetry classification followed by approximate fitting approach).  For example, see Olver \cite{FelsOlver98}.  
We describe an algorithmic approach, which appears to be new
in our paper \cite{SDGR:23}.

Difficulties with this approach are discussed, including its high computational cost, as the number of classification cases, rises dramatically with increasing dimension, and motivates and enables us to interpret the more thoroughly approximate approach given below.

Suppose we have an approximate differential equation of interest $R$, in a neighborhood containing a point $(x_0, u_0)$ in the space of independent and dependent variables of $R$.  A fundamental new contribution we make, is to define approximate Lie symmetry of $R$, in terms of exact symmetry of a nearby exact differential equation $\tilde{R}$ on a neighborhood of $(x_0, u_0)$.  If the full Lie symmetry algebra classification is available then the methods of the previous paragraph can be applied.  As we mention this method becomes impractical in higher dimensions, but establishes the existence of such nearby $\tilde{R}$, which is useful in the discussion that follows.

We say that $R$ is approximately invariant under a Lie symmetry algebra $\tilde{L}$, if there is a nearby differential equation $\tilde{R}$  that is exactly invariant under a Lie algebra $\tilde{L}$.  The practical problem, is to find ways of determining the existence of such nearby $\tilde{R}$ with large Lie symmetry algebras.  We are also interested in determining the extent of the region $S[x_0,u_0]$.

\section{Completion of Lie symmetry defining systems to involutive form}
\label{sec:chap4: completion}
The key idea in our algorithm is based on the termination of the Cartan-Kuranishi algorithm \cite{Sei10:Inv}.

Any linear homogeneous differential system $R$ with constant coefficients and differential order $q$ can be written in matrix form, with a corresponding matrix $A^{q}$, as follows:

\begin{equation}
A^{q}(x)  \underset{q}{v}  = \textbf{0}, \quad  \underset{q}{v}  = \left(
\begin{array}{c}
\underset{q}{u}\\
\vdots\\
\underset{1}{u}\\
u
\end{array}
\right)
\end{equation}
where $x = (x^1,\dots,x^n)$ and $\underset{q}{v}$ is a column vector consisting of all partial derivatives of order $\leqslant q$. The singular value decomposition (SVD) technique is used to compute the null spaces of $A^{q}(x_0)$ evaluated at $x = x_0$.

The sequence of linear homogeneous matrix systems:

\begin{equation}
A^{q}(x) \underset{q}{v}  = \textbf{0}, \quad A^{q+1}(x) \underset{q+1}{v}  = \textbf{0}, \quad A^{q+2}(x) \underset{q+2}{v} = \textbf{0}, \quad \cdots
\end{equation}
yields the prolongations of the system $R$, i.e., $DR$, $D^{2}R$, $\dots$, respectively. The right-hand side of each system is a zero vector of appropriate dimension.

Since $\text{rank} \, A = \text{rank} \, A(x_{0})$ for any generic point $x_{0} \in \mathbb{F}^{n}$, a random point $x = x_{0}$ is chosen to avoid degenerate ranks in the matrix systems. This yields a sequence of constant matrix systems:

\begin{equation}
A^{q}(x_{0}) \underset{q}{v}  = \textbf{0}, \quad A^{q+1}(x_{0}) \underset{q+1}{v}  = \textbf{0}, \quad A^{q+2}(x_{0}) \underset{q+2}{v} = \textbf{0}, \quad \cdots
\end{equation}

In the next step, the projected systems $\pi^{r}D^{k}R$, where $r = 0, 1, \dots, k$, are constructed for each prolongation $D^{k}R$. The projection $\pi$ on the vector $\underset{q}{v} \in J^{q}$, denoted $\pi \underset{q}{v} \in J^{q-l}$, is obtained by removing the coordinates of $\underset{q}{v}$ corresponding to derivatives of order $q$.

\begin{equation}
\pi^{r}R = \left\lbrace \pi^{r} \underset{q}{v} \in J^{q-l} : \quad A^{q}(x_{0}) \underset{q}{v}  = \textbf{0} \right\rbrace
\end{equation}

The key tool in this approach is the singular value decomposition. After each symbolic prolongation, the SVD of the associated matrix is computed at a random point $x_0$:

\[
A^{q+k}(x_{0}) = U \Sigma V^{\top}
\]
where $U$ is a unitary matrix, $V^{\top}$ is an orthogonal matrix, and $\Sigma$ is a diagonal matrix with diagonal entries known as the singular values. The number of nonzero singular values represents the numerical rank of $A^{q+k}(x_{0})$. A basis for the null space (kernel) of $A^{q+k}(x_{0})$ can be obtained from a sub-matrix of $V^{\top}$ by removing the first $r$ rows. This provides an estimate for $\dim (D^{k}R)$. A basis for the row space (co-kernel) is also easily computed, along with its dimension (rank).

An approximate basis for the kernel of the prolonged matrix can be obtained by considering a user-defined tolerance and setting singular values below this tolerance to zero. For the projection step, removing coordinates corresponding to the highest-order derivatives from the basis of $D^{k}R$ produces a spanning set for $\pi D^{k}R$, which can then be converted to an orthogonal basis using singular value decomposition again.


\section{Determining the structure coefficients of Lie symmetry algebras}
\label{sec:chap4: structure}

To determine the structure of Lie symmetry algebras, we use an algorithm based on Lisle, Huang and Reid's work \cite{LisHG:Sym}.

Assume that the dimension $r$ of the Lie symmetry algebra of the given DE has been identified and is finite. To fully describe this Lie algebra, we need to compute its structure constants $c^k_{ij}$ with respect to a suitable basis. In this section, we outline how to numerically approximate the structure constants $c^k_{ij}$ using the defining system $\Rdef$ of the Lie symmetry algebra.

By the Cartan-Kuranishi prolongation theorem \cite{CarKura}, and the equivalence of Cartan involutivity with projective involutivity \cite{BLRZ}, there exists a finite prolongation $D^{k'}\Rdef$ and an integer $\ell$ (with $0 \leq \ell \leq k$) such that $\pi^{\ell}D^{k'}\Rdef$ is involutive. Assume that $D^{k'}\Rdef$ has an associated coefficient matrix $A^{(q+k')}(z_0)$, evaluated at a random point $z = z_0$. Since the system $\pi^{\ell}D^{k'}\Rdef$ is involutive, data up to order $q' = q+k'-\ell$ can be extended uniquely to local solutions at $z_0$. The core idea is to uniquely determine the Lie bracket and the structure constants using the Cartan-Kuranishi existence and uniqueness theorem.

To determine the structure constants, we need derivatives up to order $q'+1$. This ensures that the system remains uniquely determined by neighboring involutive systems. Without loss of generality, assume both $\pi^{\ell}D^{k'}\Rdef$ and $\pi^{\ell-1}D^{k'}\Rdef$ are involutive. If not, we use their prolongations: $\pi^{\ell}D^{k'+1}\Rdef = D\pi^{\ell}D^{k'}\Rdef$ and $\pi^{\ell-1}D^{k'+1}\Rdef = \pi^{\ell}D^{k'}\Rdef$, which would ensure involutivity at the $(k'+1)$-th prolongation.

Given an input tolerance $\tol > 0$, we compute the SVD at the random point $z_0$:
\[
A^{(q+k')}(z_0) = U \Sigma V^T
\]
Singular values smaller than $\tol$ are replaced by zeros, resulting in a nearby matrix:
\[
A' = U \Sigma' V^T = \begin{bmatrix} U_1 & U_0 \end{bmatrix} \begin{bmatrix} \Sigma_1 & 0 \\ 0 & 0 \end{bmatrix} \begin{bmatrix} V_1^T \\ V_0^T \end{bmatrix}
\]
where the diagonal matrix $\Sigma_1$ contains singular values larger than $\tol$ \cite{CarKura}. The columns of $V_0$ form an orthogonal basis for the numerical null space of $A^{(q+k')}(z_0)$. We project these vectors to obtain $\pi^{\ell}V_0 = \{\pi^{\ell}(v) : v \in V_0\}$ by removing coordinates corresponding to derivatives of order $> q'$. This provides a spanning set for the involutive system $\pi^{\ell}D^{k'}\Rdef$ at $z = z_0$.

To compute the structure constants, we need a basis for the space spanned by the columns of $\pi^{\ell}V_0$. This is achieved by applying the SVD again to this matrix, and replacing singular values below $\tol$ with zeros. The first $r$ columns of the resulting $U$ matrix form an orthogonal basis $B^{(q')}$ for the approximate column space of $\pi^{\ell}V_0$. Since the system is of finite type, and both $\pi^{\ell}D^{k'}\Rdef$ and $\pi^{\ell-1}D^{k'}\Rdef$ are involutive, this basis $B^{(q')}$ can be uniquely extended to a basis $B^{(q'+1)}$ for the column space of $\pi^{\ell-1}V_0$ by solving for the coordinates of order $q'+1$.

Each basis vector $v^{(q'+1)}_i$ (from $B^{(q'+1)}$) uniquely determines a local symmetry vector field $X_i = \zeta^j_i \partial / \partial z^j$. The vector fields close under the Lie bracket, giving the structure constants $c^k_{ij}$ as:
\[
c^k_{ij} = \langle v^{(q')}_k, w^{(q')}_{ij} \rangle
\]
where $w^{(q')}_{ij}$ represents the components of the Lie bracket for the vector fields $X_i$ and $X_j$.

\section{Reliability of the approximate Lie algebra results}
\label{sec:chap4: reliability}

Suppose A is a $m$x$n$ matrix, $A=U\Sigma V^T \rightarrow AV = U\Sigma$. Since we apply SVD to our simplified prolonged approximate system if we compute $r=rank(A)$, we will have $V(:,1:r)$ spans the row space, which is the basis of row space, and $v(:,r+1:n)$ spans the null space, which is the basis of the kernel, or, the basis of null space. Using Lisle's method in previous section \cite{LisHG:Sym} we can calculate the commutator numerically using the basis of null space derived by SVD. The structure constants $c_{ij}^k$, which are given by


\begin{equation}
\label{eq4:cijk}
	[L_i,L_j] = \sum_{k=1}^r c_{ij}^k L_k, \quad 1 \leq i, j \leq d
\end{equation}

and calculated by

\begin{equation}
    \label{eq4:cijkwij}
        c_{ij}^k = w_{ij}^{(q'-1)}(V_k^{q'-1)})^T
\end{equation}

where $ w_{ij}^{(q')} = \sum_k c_{ij}^k V_k^{(q')}$. Here $q'$ is the involutive order we achieved after the approximate differential-elimination algorithm(GIF).

\begin{figure}[ht]
\centering
		\includegraphics[angle=0,origin=c, width=0.5\textwidth]{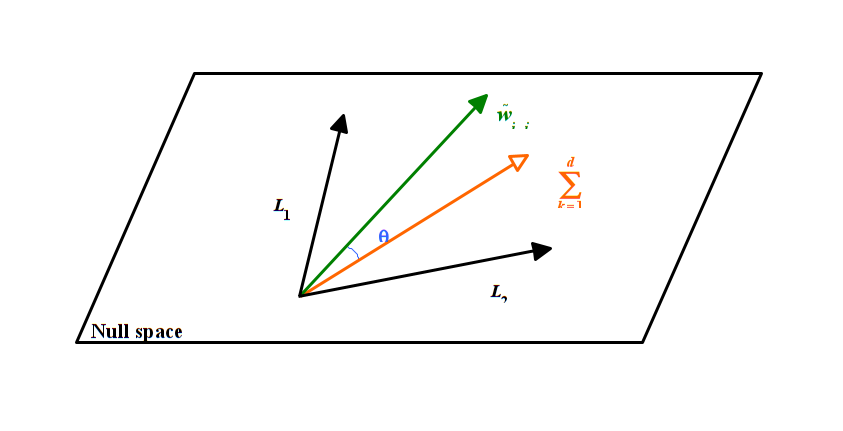}
		\caption{The commutator computed from SVD may not be on the null space}
\label{fig4:Commutator}
\end{figure}

However, how reliable is the method?  When a nearby candidate for a Lie algebra is found, is it really close to a Lie algebra?  The main problem we want to focus on is after calculation involving floating point numbers (obviously due to SVD), the numerical commutator might not be on the null space exactly.  From \ref{fig4:Commutator}, there will be a small angle $\theta$ between the commutator and the null space, which shows the approximate commutator is not on the null space, because of some error (normal for numerical computation).

Then, to consider the quality of approximate symmetries, we use the absolute error

\begin{equation}
    \label{eq4:sigmaw}
    \left\|w_{ij}-\tilde{w}_{ij}\right\| = \left\|\sum_k c_{ij}^k V_k-\tilde{w}_{ij}\right\| 
\end{equation}
where $w_{ij}$ is a linear combination of $V_k$ which should be on the null space. And $\tilde{w}_{ij}$ is directly calculated by Lisle's method so it is approximately computed and may not be on the null space exactly. If the absolute error is very close to 0, or it is smaller than the tolerance we give, we say that the quality of approximate symmetries is good. 


A better measure of error is the relative error

\begin{equation}
\label{eq4:rew}
\sigma_{\mbox{Lie}}(w)=\frac{\left\|\sum_k c_{ij}^k V_k-\tilde{w}_{ij}\right\| }{\left\|\sum_k c_{ij}^k V_k\right\|}
\end{equation}



Another way to consider the reliability is to calculate the Jacobi identity. See \cite{hall2015lie}. Jacobi identity of Lie algebras is given by:

\begin{equation}
\label{sigmajacobi}
    [L_i,[L_j,L_k]]+[L_j,[L_k,L_i]]+[L_k,[L_i,L_j]]=0
\end{equation}.

We similarly define $\sigma_{\mbox{Jacobi}}$ via the SVD by using a suitable norm of \ref{sigmajacobi}.

\section{Preliminary experiment results}
\label{sec:chap4: Example}

Consider a second order ODE:

\begin{equation}
    \label{eq5: ODEex}
    \frac{d^2}{dx^2}u(x) + 0.708203932 u(x)\frac{d}{dx}u(x) + 5u(x)^3 + e^{-10((x-1)^2+(u(x)-1)^2)}=0.
\end{equation}

\begin{figure}
    \centering
    \includegraphics[width=\textwidth]{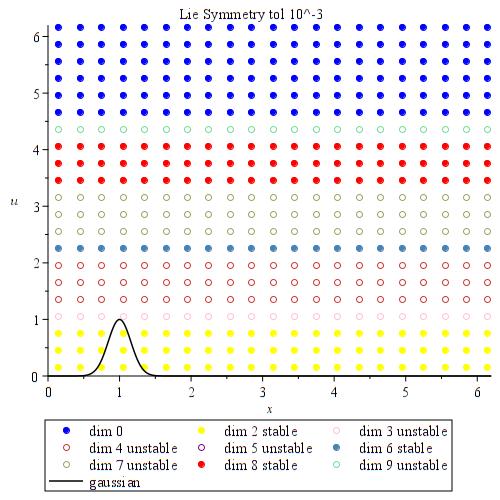}
    \caption{Dimension plot with tolerance $10^{-3}$. $x=0..5$, $u=0..5$} 
    \label{fig5: tol3}
\end{figure}

\begin{figure}
    \centering
    \includegraphics[width=\textwidth]{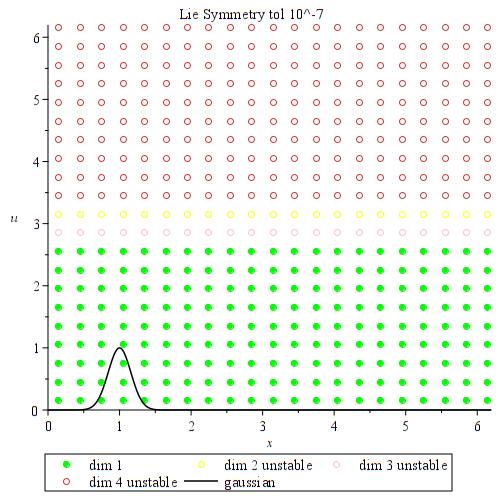}
    \caption{Dimension plot with tolerance $10^{-7}$. $x=0..5$, $u=0..5$} 
    \label{fig5: tol7}
\end{figure}

\begin{figure}
    \centering
    \includegraphics[width=\textwidth]{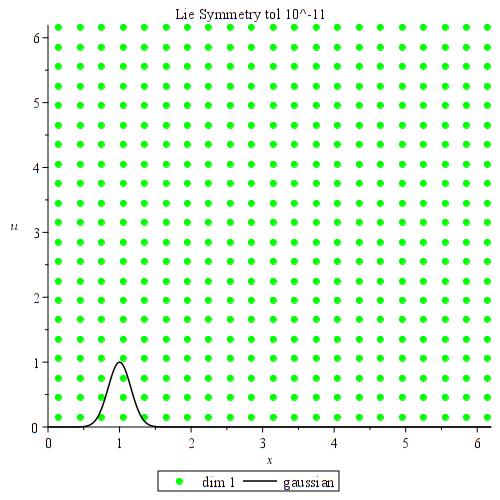}
    \caption{Dimension plot with tolerance $10^{-11}$. $x=0..5$, $u=0..5$} 
    \label{fig5: tol11}
\end{figure}

By applying our algorithms, the approximate Lie symmetry and its quality were calculated. The result is shown in the figure \ref{fig5: tol3}, \ref{fig5: tol7}, \ref{fig5: tol11}.

From \ref{fig5: tol3}, which we give a large tolerance, we can find different dimension regions directly. We put the $e^{-10((x-1)^2+(y(x)-1)^2)}$ as a Gaussian function plot on the graph, and clearly, we can see that points near the function are 2-dimensional, which is the exact symmetry of (\ref{eq5: ODEex}) without Gaussian function. Then we have a solid 6-dimensional region and an 8-dimensional region, where in between is an unstable region, which is the transition area mentioned before. When the point is far enough from the Gaussian function, it returns to 0-dimensional because we have a $y(x)^3$ term, which has a great impact when $y(x)$ is large. However, given such a large tolerance, even we get good quality approximate symmetries, we cannot conclude the nearby system $\tilde{R}$ with the same $\tilde{L}$ are close enough to $R$. When the tolerance becomes smaller, such as $10^{-7}$ in \ref{fig5: tol7}, we only have a 1-dimensional stable area, which is the exact symmetry of \ref{eq5: ODEex}, where above the solid region we have the transition area. When the tolerance becomes very small (such as $10^{-11}$ in this case), more points will have the same approximate Lie symmetry as the exact symmetry derived by \maple{rifsimp}. See \ref{fig5: tol11}.

\section{Extending the Regions of Validity for an Approximate Lie Algebra}
\label{sec:chap4: region}

Given the point $(x_0, u_0)$ in the exact case at all (non-singular) point we expect one parameter Lie algebra though the structure constants will vary because the algorithm for their determination depends on the point $(x_0,u_0)$.  However there will exist a change of basis, under which these different representations will be mapped to each other.

But in the case of approximate Lie symmetry algebras, by our approach, there will be a neighborhood of each point having the same associated approximate Lie algebra structure, and different points may have non-isomorphic Lie algebras (i.e. their structure constants are not related by a change of basis).  Moreover there are also regions in which no Lie algebra structure, could be reliably determined.

It is natural to extend the algorithm described in the previous sections for computing Lie symmetry algebras at points $(x_0, u_0)$ to a grid of such points, while partitioning with respect to Lie Algebra isomorphism.

Let $L =$ span$\{x_1, \cdots , x_n\}$ and $L^* =$ span$\{u_1, \cdots , u_n\}$ be two n-dimensional $\mathbb{F}$-Lie algebras
with structure constants $a^k_{ij}$ and $b^k_{ij}$, respectively.
Given a tolerance $\tol$, define the equivalence relation $P \sim P^*$ with $(x_0, u_0) \sim (x_0^*, u_0^*)$ iff $dim(L)=dim(L^*)$ and there is a change of basis taking $L$ into $L^*$.

Assume that the matrix [$\phi$] of $\phi$ is as follows:

\begin{equation}
    [\phi]=\begin{bmatrix}
    z_{11} &\cdots &z_{n1} \\
    \vdots & \ddots & \vdots \\
    z_{1n} &\cdots &z_{nn}
    \end{bmatrix}
\end{equation}.

We restrict ourselves to the case of finite dimensional Lie algebras, and the infinite dimensional case (Lie groupoids) is an interesting open problem.

Note that numerically we test the existence of the Lie algebra isomorphism by using the equations between structure constants for $L$ and $L^*$ given by Nguyen, Le and Vo \cite{TVV21}:

\begin{equation}
    \label{eq5: iso}
    \left\{
    \begin{aligned}
    &\sum_{k=1}^{n} z_{ks} a_{ij}^{k} - \sum_{k=1}^{n} \sum_{l=1}^{n} z_{ik} z_{jl} b_{kl}^{s} = 0, \quad 1 \leq i < j \leq n, \quad s = 1, \dots, n, \\
    &1 - z \det[\phi] = 0 \quad (z\in \mathbb{F}\quad \text{as a new unknown})
    \end{aligned}
    \right.
    \end{equation}
which consists of $\leq n\binom{n}{2}+1$ polynomials in $\mathbb{F}[z,z_{11},\cdots, z_{1n},\cdots, z_{n1},\cdots, z_{nn}]$ with $n^2+1$ unknowns $z,z_{ij} \in \mathbb{F}$.

The first equation in (\ref{eq5: iso}) is equivalent to $\phi[x_i,x_j] = [\phi[xi], \phi[x_j]]$, and the second equation in (\ref{eq5: iso}) has the same meaning as $\text{det}[\phi] \neq 0$ when using exact triangular decomposition, as used by those researchers. However, our case has approximate structure constants, so numerical methods are needed.
We will use the methods of numerical algebraic geometry while avoiding the expensive determinant term for this task.

\textbf{Acknowledgment:} We thank Peter Olver, who suggested using the angle instead of absolute error.

\end{document}